\documentclass[prd,aps,twocolumn,aps,amsmath,amssymb,nofootinbib,preprintnumbers]
{revtex4}
\usepackage[dvips]{graphicx}
\usepackage{epsf}

\usepackage{graphicx}% Include figure files
\usepackage{dcolumn}% Align table columns on decimal point
\usepackage{bm}% bold math
\usepackage{amsmath}
\usepackage{amsfonts}
\usepackage{subfigure}
\def\ls{\mathrel{\lower4pt\vbox{\lineskip=0pt\baselineskip=0pt
           \hbox{$<$}\hbox{$\sim$}}}}
\def\gs{\mathrel{\lower4pt\vbox{\lineskip=0pt\baselineskip=0pt
           \hbox{$>$}\hbox{$\sim$}}}}
\def\drawbox#1#2{\hrule height#2pt

\hbox{\vrule width#2pt height#1pt \kern#1pt
              \vrule width#2pt}
              \hrule height#2pt}

\def\Asym#1#2{\vcenter{\vbox{\drawbox{#1}{#2}
              \kern-#2pt       % line up boxes
              \drawbox{#1}{#2}}}}

\newcommand{\e}{\textrm{e}}

\def\be{\begin{equation}}
\def\beq\begin{equation}
\def\ee{\end{equation}}
\def\bea{\begin{eqnarray}}
\def\eea{\end{eqnarray}}

\def\beq{\begin{equation}}
\def\eeq{\end{equation}}
\def\beqa{\begin{eqnarray}}
\def\eeqa{\end{eqnarray}}

\def\cR{{\mathcal R}}

\newcommand{\bmat}{\left(\begin{array}}
\newcommand{\emat}{\end{array}\right)}
\newcommand{\w}{\omega}

\begin{document}

\title{Large tensor-to-scalar ratio and low scale inflation}

\author{Rouzbeh Allahverdi$^{1}$}
\author{Anupam Mazumdar$^{2,3}$}
\author{Tuomas Multam\"aki$^{4}$}

\affiliation{$^{1}$~Department of Physics and Astronomy, University of New Mexico,
Albuquerque, NM 87131, USA\\
$^{2}$~Physics Department, Lancaster University, Lancaster, LA1 4YB, UK\\
$^{3}$~Niels Bohr Institute, Blegdamsvej-17, Copenhagen-2100, Denmark\\
$^{4}$~Department of Physics, University of Turku, FIN-20014, Finland}

%\date{May 3, 2006}

\begin{abstract}
It is plausible that the scalar density perturbations are
created by a relatively low scale model of inflation which matches
the observations of CMB anisotropy and excites Standard Model baryons and cold dark matter,
but generates negligible gravity waves. Nevertheless a significantly large tensor perturbations
can be observed if there exists a prior phase of high scale inflation
separated by a matter or radiation dominated epoch. In this paper we provide a simple example where
gravity waves generated at high scales can trickle through
the horizon of the second phase of inflation and leave a distinct imprint in the
spectrum of the tensor modes with a strong red tilt. A first phase of
{\it assisted inflation} occurring at a high scale $H \sim 10^{13}$~GeV is followed by a second
phase of {\it MSSM inflation} which happens at $H \sim 1$~GeV. The largest tensor-to-scalar
ratio is then bounded by $r_{\rm observed} \leq 0.8$ on the largest scales,
roughly of the size of the horizon.
\end{abstract}

\maketitle

%%%%%%%%%%%%%%%%%%%%%%%%%%%%%%%%%%%%%%%%%%%%%%%%%%%%%%%%%%%%%%%%%

The positive detection of stochastic gravitational waves, especially
at very large angular scales, is considered to be a strong support for
the inflation paradigm. It is believed that the PLANCK satellite will be
able to detect gravity waves if the tensor to scalar ratio is significant, i.e. $r \equiv {\cal T}/{\cal S}\sim 0.01-0.1$~\cite{PLANCK}. It is however a daunting task to generate
large tensor perturbations in a reasonable model of inflation which is embedded
within particle physics~\cite{anu} and in string inspired models of
inflation~\cite{Kallosh}. In realistic models based on the minimal
supersymmetric standard model (MSSM)~\cite{AEGM,AEGJM,AKM}, where the
inflaton carries the standard model (SM) charges and eventually decays
into the SM baryons and the cold dark matter, inflation occurs at
very low scales and generated gravity waves are too small to be detectable in
the future experiments.

A significant tensor to scalar ratio is obtained in chaotic
inflation~\cite{Lindebook}. However in this class of models
the vacuum expectation value (VEV) of the inflaton exceeds the
Planck scale. Moreover, the inflaton is an absolute gauge singlet
carrying no SM charges, thus rendering the model with no real
connection to particle physics. A plausible scenario with sub-Planckian
VEVs, called {\it assisted inflation}, arises when multi-scalar fields
collectively drive inflation~\cite{assist1,assist2}. Assisted inflation has found more
realistic connection to particle physics~\cite{JM} and string
theory~\cite{assist3,Ninf}. It can generate a reasonable
tensor-to-scalar ratio which could potentially be detected by the future
CMB experiments.

In spite of these successes, assisted inflation can not automatically
explain the SM baryons and cold dark matter. The model can at
best be embedded in a hidden sector, so far, whose couplings to the SM fields are essentially set by hand. This situation remains
in string theory, as the inflation sector and the matter sector
are often not the same~\cite{Myers}.

Tensor perturbations in two field inflationary models
have been studied in~\cite{Polarski}, where both the phases of inflation occur
at super-Planckian VEVs. It was envisaged that a contribution to the
scalar perturbations came from both phases of inflation, thus producing a
break in the power spectrum for the temperature anisotropy. Similar break
in power spectrum for tensor perturbations were studied in~\cite{Brandenberger,Mukhanov}.

In this paper our goal is to avoid inflation at super-Planckian VEVs while still generating significant tensor perturbations with a large tensor-to-scalar ratio. Secondly, we wish to generate the entire scalar perturbations during the second phase of inflation in order not to have any observable break in the power spectrum for the CMB temperature anisotropy. Third important point is that we wish to present a realistic model of inflation where reheating to the Standard Model baryons and cold dark matter is automatic and there are no uncertainties in the inflaton coupling to matter.

Here we aim to keep all of the virtues of an MSSM based
low scale inflation with direct connection to rich phenomenology~\cite{AEGM,AEGJM,AKM},
but modify the physics at higher energies, such that gravity waves can trickle past
through the horizon of the MSSM inflation. The idea will be to have
two phases of inflation where the first one contributes to the tensor modes,
while the latter provides scalar density perturbations and successful
reheating into the (dark) matter. We strictly assume
that the entire CMB temperature anisotropy arises from the second phase of inflation, therefore,
there is no break in its power spectrum. Moreover, the
first phase of inflation can also set initial conditions for a low scale inflation as we will
discuss briefly

\begin{figure}
\includegraphics[width=5cm]{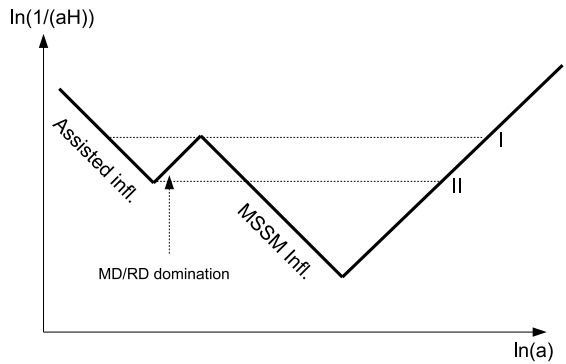}
\caption{Schematic diagram of the two phases of inflation separated by
matter/radiation dominated epoch. Mode 1 skims through the horizon
of the second phase of inflation. Note that assisted inflation happens at a high
scale while MSSM inflation takes place at low scales.}\label{infladiagram}
\end{figure}

In Fig. \ref{infladiagram} we show a schematic
diagram of the scenario, two inflationary phases separated by a
matter/radiation dominated epoch. For the purpose of our discussion,
without loss of generality, we will only concentrate on a matter dominated
intermediate epoch. The largest observable scale is somewhere between
the scale where the first phase of inflation, mode I, ends and the second
starts, mode II. Here for simplicity we mainly concentrate on the case
where the largest observable scale corresponds to the scale where the
MSSM inflation starts, i. e. mode I in the figure.

%%%%%%%%%%%%%%%%%%%%%%%%%%%%%%%%

%%%%%%%%%%%%%%%%%%%%%%%%%%%%%%%%%%

There will be a clear prediction for the tensor modes in our case
compared to the case where both of the tensor and scalar perturbations are
generated by a single phase of inflation. The distinction is that we
would see the tensor modes {\it only} on the largest angular
scales, i.e. $l=2$, and there will be a sharp drop in the correlations involving
the tensor component at smaller angular scales with a power dropping $\sim k^{-5}$.
We will also see that scalar and tensor modes arising from the first
phase of inflation are anti-correlated.

%%%%%%%%%%%%%%%%%%%%%%%%%%

\section{Highlights of assisted inflation and gravity waves}

In order to illustrate our case, let us assume that a first phase of
assisted inflation is driven by $n$ scalar fields with an identical potential:
\begin{equation}\label{1}
V=\sum_{i} m^2\chi_{i}^2\,.
\end{equation}
This is well studied in Ref.~\cite{Kim}. The scalar and tensor
power spectrum, denoted by ${\cal P}^{\rm assist}_{R}$ and
${\cal P}^{\rm assist}_g$ respectively, are given by:
\begin{eqnarray} \label{r1}
{\cal P}^{\rm assist}_{\cal R} & \approx & \frac{H^2\sum_{i}\chi_{i}^2}{16 \pi^2M_{P}^4}\, , \\
\label{r2}
{\cal P}^{\rm assist}_{g}& = & \frac{2H^2}{\pi^2 M_{P}^2}\, , \\
\label{r3}
r_{\rm assist}& = &\frac{{\cal P}^{\rm assist}_{g}}{{\cal P}^{\rm assist}_{\cal R}}
\approx \frac{8}{\cal N}\, ,
\end{eqnarray}
where $r_{\rm assist}$ is the tensor-to-scalar ratio and ${\cal N}$
is the number of e-foldings between the time the relevant modes exit the
horizon and the end of inflation.

The observed value of ${\cal P}_{\cal R}$ is set by COBE normalization, i.e.
${\cal P}_{\cal R} \simeq 6 \times 10^{-11}$. Hence detectable gravity waves, i.e. $r \geq 0.1$,
can be generated if~\footnote{For the purpose of illustration we will assume that
${\cal P}_{\cal R}^{\rm assist} \leq 6 \times  10^{-11}$.}:
\begin{equation} \label{assist}
H_{\rm assist} \simeq 10^{13}~{\rm GeV}\,.
\end{equation}
After inflation the $\chi$ fields will start oscillating coherently (we
assume that they have the same amplitude of oscillations
by virtue of identical mass scale). There will be a long wait until
the next phase of inflation begins because MSSM inflation occurs
at a low scale, i.e. $H_{\rm MSSM} \sim 1$~GeV~\cite{AEGM}.
Since all $\chi$ fields are gauge singlets, it is conceivable that
their couplings to matter are extremely weak, and hence they
do not decay for $H > 1$ GeV~\footnote{The gauge singlet inflaton
couples to the SM fermions and gauge fields via dimension-$5$ operators,
i.e. non-renormalizable interactions~\cite{Jaikumar}, leading
to a small decay rate. It can have a renormalizable coupling to the SM Higgs,
but for an absolute gauge singlet the strength of such a coupling is not
constrained by any symmetry argument. Moreover, such a coupling
does not necessarily lead to a complete decay of the inflaton.}.
Then the universe will evolve as in a matter dominated epoch for
$H_{\rm MSSM} < H < H_{\rm assist}$.

During this intermediate epoch modes which exit the horizon
in the last ${\cal N}_{\rm re-enter}$ e-foldings of assisted inflation,
i.e. between mode I and II in Fig. 1, will re-enter
the horizon set by the MSSM inflation~\cite{minflation}:
\be \label{efold}
{\cal N}_{\rm re-enter} = \left(\frac{1-n}{2}\right) \ln \left(\frac{V_{\rm assist}}
{V_{\rm MSSM}}\right) \simeq 10 \,.
\ee
For $H_{\rm assist} \sim 10^{13}$~GeV and $H_{\rm MSSM} \sim 1$~GeV
and $n=2/3$ for matter domination the relevant mode which leaves before the
end of assisted inflation horizon leads to ${\cal N}_{\rm re-enter} \simeq 10$
e-foldings~\footnote{For radiation domination  we have $n=1/2$, which
results in ${\cal N}_{\rm re-enter} \simeq 15$.}. For example, the
mode I in Fig.~(1) leaves $10$ e-foldings before the end of assisted inflation.

The crucial observation of this paper is that the tensor-to-scalar
ratio in such a two phase inflationary model, where the tensor perturbations from the first phase
skim through the horizon of the second phase of inflation (i.e. mode 1 in Fig.~1),
is always bounded by
\begin{eqnarray}\label{imp}
r_{\rm observed}  &\leq & \frac{8}{{\cal N}_{\rm re-enter}}\left(\frac{{\cal P}^{\rm assist}_{\cal R}}
{{\cal P}^{\rm MSSM}_{\cal R}}\right)\,, \nonumber \\
&\leq & 0.8\left(\frac{{\cal P}^{\rm assist}_{\cal R}}
{{\cal P}^{\rm MSSM}_{\cal R}}\right )\,.
\end{eqnarray}
This assumes that the entire scalar perturbations are generated by the second
phase of inflation, but its contribution to tensor perturbations
is negligible. This result is generic and does not depend on the details
of either the first phase or the second phase of inflation. The numerical value
only depends on the Hubble expansion rates for the two phases. In the most optimistic scenario
when both the phases of inflations generate similar amplitude, spectral tilt
and the running of the spectral tilt, the above bound can be saturated.
On the other hand, if the amplitude of scalar perturbations from the second phase
is dominant, then the bound is strict, $r_{\rm observed} < 0.8$.

Another important comment is that if there was a single phase of inflation driven by
the assisted inflation alone, then the tensor-to-scalar ratio would be determined by
Eq.~(\ref{r3}). In this case the relevant scalar perturbations are produced
at roughly ${\cal N} \sim 60$, yielding $r_{\rm assist} \sim 0.13$, which is
lower than the maximum value allowed in our scenario, see Eq.~(\ref{imp}).

%%%%%%%%%%%%%%%%%%%%%%%%%%%%%%%%%%

\section{Highlights of MSSM inflation and scalar perturbations}

We now briefly discuss MSSM inflation (for details, see~\cite{AEGM,AEGJM,AKM}).
Inflation is driven by MSSM flat directions, which are classified by {\it gauge-invariant}
combination of scalar fields in the theory. The potential along such a
flat direction (after being minimized along the angular direction) is given by:
\beq \label{scpot}
V = {1 \over 2} m^2_\phi\ \phi^2 - A
{\lambda_{n}\phi^n \over n\,M^{n-3}_{\rm P}} + \lambda^2_n
{{\phi}^{2(n-1)} \over M^{2(n-3)}_{\rm P}}\,,
\eeq
Here $\phi$ denotes the radial component of the flat direction field, and
$A$ is a positive definite quantity of dimension mass. If $A$ and $m_\phi$ are
related by $A^2 = 8 (n-1) m^2_\phi$, there is a saddle point:
\beq \label{phi0}
\phi_0 = \left({m_\phi M^{n-3}_{\rm P}\over
\lambda_n\sqrt{2n-2}}\right)^{1/(n-2)}\,.
\eeq
where $V^{\prime}(\phi_0) = V^{\prime \prime}(\phi_0)=0$. The potential
is very flat near $\phi_0$, and it is given by:
\beq \label{potential}
V_{\rm MSSM} = {(n-2)^2\over2n(n-1)}\,m^2_\phi \phi_0^2\,.
\eeq
As a result, if the flat direction field is in the vicinity of $\phi_0$ (and
has a sufficiently small kinetic energy), there will be an ensuing
phase of inflation. The Hubble expansion rate during inflation is
given by
\beq \label{hubble}
H_{\rm MSSM} = {(n-2) \over \sqrt{6 n (n-1)}}
{m_{\phi} \phi_0 \over M_{\rm P}}\,.
\eeq
Inflation ends when $\vert \eta_{inf} \vert \sim 1$, where $\epsilon \equiv
(M_{\rm P}^2/2)(V^{\prime}/V)^2$ and $\eta_{inf} \equiv M^2_{\rm
P}(V^{\prime \prime}/V)$ are the slow roll parameters. The number of
e-foldings between the time when the observationally relevant
perturbations are generated and the end of inflation
follows~\cite{minflation,andrew}:
\be
{\cal N}_{\rm COBE} \simeq 66.9 + {1 \over 4} ~ {\rm ln} \Big({V_{\rm MSSM} \over M^4_{\rm P}}\Big) \simeq 50.
\ee
Here we have used the fact that, due to efficient reheating~\cite{AEGJM}, the
energy density in the inflaton gets converted into radiation very
quickly after the end of MSSM inflation. The amplitude of scalar
perturbations thus produced is given by:
\beq \label{ampl}
\delta_{H} \simeq
\frac{1}{5\pi} \sqrt{\frac{2}{3}n(n-1)}(n-2) ~ \Big({m_\phi M_{\rm P} \over
\phi_0^2}\Big) ~ {\cal N}_{\rm COBE}^2\,.
\eeq
We remind that the scalar power spectrum ${\cal P}_{\cal R}$
is related to $\delta_H$ through ${\cal P}_{\cal R} = (4/25) \delta^2_H$.

MSSM inflation can
produce up to $10^{3}$ e-foldings of slow roll inflation~\cite{AEGM,AEGJM}. However 
the exact number of e-foldings in the slow roll regime depends on
the initial VEV of the inflaton. Since in our case assisted inflation sets
the initial conditions for the MSSM inflation, we are focusing on those
Hubble patches where the inflaton VEV matches to provide
${\cal N}_{\rm COBE}$. On the other hand, in those regions where this criteria
is not met, it would be hard to see any trace of tensor modes on the largest angular
scales. 

%The scalar spectral index is found to be within the WMAP 3-years' data
%allowed range, $0.92 \leq n_{s} \leq 1.0$ (the detailed calculation can
%be found in~\cite{Lyth,AEGJM,ADM})~\footnote{The spectral index is
%given by: $ n_s = 1 - 4 \sqrt{\Delta^2} ~ {\rm cot} [{\cal N}_{\rm
%COBE}\sqrt{\Delta^2}]$~\cite{Lyth,AEGJM,ADM}. For $0 \leq \Delta^2 \leq {\pi^2 \over 4 {\cal
%N}^2_{\rm COBE}}$, the entire allowed range $0.92 \leq n_s \leq 1.0$ is covered.}.

For weak scale supersymmetry, i.e. $m_{\phi} \simeq 100~{\rm GeV}-10~{\rm TeV}$,
acceptable $\delta_H = 1.91 \times 10^{-5}$ and $0.92 \leq n_s < 1$~\footnote{The spectral index is
given by: $ n_s = 1 - 4 \sqrt{\Delta^2} ~ {\rm cot} [{\cal N}_{\rm
COBE}\sqrt{\Delta^2}]$~\cite{Lyth,AEGJM,ADM}. For $0 \leq \Delta^2 \leq {\pi^2 \over 4 {\cal
N}^2_{\rm COBE}}$, the entire allowed range $0.92 \leq n_s \leq 1.0$ is covered.}, compatible with the WMAP 3-year's data, are obtained if $n=6$ and
$\lambda \sim 1$~\cite{AEGM}, or if $n=3$ and
$\lambda \sim 10^{-12}$~\cite{AKM}~\footnote{The inflaton is a linear
combinations of slepton or squark fields in the former case~\cite{AEGM}.
In the latter case, a linear combination of the Higgs, slepton and sneutrino
fields play the role of the inflaton~\cite{AKM}.}. Then from Eqs.~(\ref{phi0},\ref{hubble}),
we have $\phi_0 \sim 10^{14}$ GeV and $H_{\rm MSSM} \sim {\cal O}(1~{\rm GeV})$.

Another important point for an MSSM inflation is that the model
parameters naturally accommodate thermal dark matter, i.e. successful
inflation is compatible with the allowed regions of the parameter space
which yield the correct abundance of neutralino dark matter~\cite{ADM}.

Although we invoke assisted inflation for generating gravity waves, it also
serves another useful purpose as it can set the initial conditions for the
MSSM inflation~\cite{AFM}. The phase of assisted inflation will induce large
quantum fluctuations to the flat direction since $m_{\phi} \ll H_{\rm assist}$.
The flat direction will make quantum jumps in this phase which accumulate
in a random walk fashion~\cite{Lindebook}:
${d \langle \phi^2 \rangle / dt} = {H^3 / 4 \pi^2}$. Then the required
number of e-foldings  of assisted inflation, which is required to push
$\langle \phi^2 \rangle$ to the vicinity of $\phi_0$ follows
${\cal N}_{\rm assist} = \sqrt{6 \pi^2} {\phi_0 /m}$~\cite{AFM}. For
$\phi_0 \sim 10^{14}$~GeV and $m \approx 10^{13}$~GeV, the required
number of e-foldings turns out to be ${\cal N}_{\rm assist} \sim 80$. This is quite
feasible in assisted inflation with a sufficiently large number of
fields~\cite{JM,Alabidi,andrew}~\footnote{An extreme flatness of the MSSM inflaton potential
ensures that the flat direction VEV remains virtually frozen during the intermediate stages.}.

%%%%%%%%%%%%%%%%%%%%%%%%%%%%%%%%%%%%

\section{Analysis of tensor and scalar modes}

Now we wish to analyze the evolution of tensor and scalar modes from the first
phase of inflation. Note that the phase of MSSM inflation
requires at least $50$ e-foldings in order to explain the observed scalar
perturbations~\cite{AEGM,AEGJM}. We are interested in those
tensor modes which just skim through the horizon of the MSSM inflation,
i.e. modes that were of the horizon size when MSSM inflation began.
They will be observable today if MSSM inflation lasted the minimum
number of required e-foldings, i.e. ${\cal N}_{\rm COBE}={\cal N}_{\rm MSSM}  \simeq 50$.

These modes (tensor and scalar) left the horizon about
${\cal N}_{\rm re-enter} \simeq 10$ e-foldings before the
end of assisted inflation, see Eq.~(\ref{efold}). Modes that
left earlier will be well outside the horizon when MSSM
inflation started, and hence today as well. On the other hand,
modes that left later will re-enter the horizon before the onset of MSSM
inflation and get suppressed. In this Section, we will discuss this
issue in detail.

%%%%%%%%%%%%%%%%%%%%%%%%%%%%%%%%%%%%%%

\subsection{Brief summary of scalar perturbations}

The evolution of the scalar curvature perturbations in the early
universe can be followed by using the {\it gauge-invariant}
Mukhanov variable $u$~\cite{mfb}, which on flat hypersurfaces is related to
the fluctuation of the scalar field, $\delta \phi$, simply by $u=a \delta \phi$.
The curvature perturbation, $\cR$, is related to the variable $u$
by~\footnote{Although in assisted inflation there are multiple fields, during the
inflationary epoch there exists a unique late time attractor trajectory~\cite{assist1}. This
ensures that the perturbations are determined solely by the Mukhanov variable
corresponding to a single field, therefore, there is only adiabatic mode and no
isocurvature (or entropy) mode, see ~\cite{assist1,Kim}. This is particularly true in our case since the masses
of all the fields are the same.}
\be\label{udef}
u=-z\cR,
\ee
where the variable $z$ is defined as $z\equiv a\dot{\phi}/H$.
The Fourier modes of $u$, denoted by $u_k$, evolve according to \cite{mfb}
\be \label{uequ}
\partial_\tau^2u_k + \left( k^2 - \frac{\partial_\tau^2 z}{z}
\right)u_k=0,
\ee
where 
%$u_k$ is the Fourier component of Mukhanov variable and
$\tau$ denotes conformal time. The power spectrum of the
curvature perturbation is given by
\be\label{power}
\mathcal{P}_{\mathcal{R}}^{1/2}(k)=\sqrt{\frac{k^3}{2\pi^2}}
\left| \frac{u_k}{z}\right|.
\ee
Deep inside the Hubble radius, when $k\gg (\partial_\tau ^2z)/z$,
the solution of Eq. ~(\ref{uequ}) tends to the free field
Bunch-Davies vacuum solution (with arbitrary normalization)
\be\label{freesol}
u_k\propto \frac{1}{\sqrt{2k}}\e^{-ik\tau}.
\ee
Well outside the Hubble radius, $k\ll (\partial_\tau ^2z)/z$,
the growing solution is simply
\be\label{freesol2}
u_k\propto z,
\ee
indicating that the curvature perturbation tends to a constant outside
the Hubble radius. From these considerations the qualitative
picture is clear: the vacuum modes are stretched during inflation
until they exit the horizon, resulting in a constant
curvature perturbation at superHubble scales.

%%%%%%%%%%%%%%%%%%%%%%%%%%%%%%%

\subsection{Brief summary of tensor perturbations}

For the tensor modes, the situation is similar to the scalar case but
there are also important differences. The equation determining the evolution
of modes is now
\be\label{gravmodes}
v''_k+\left(k^2-\frac{a''}{a}\right)=0,
\ee
where $v_k$ is related the gravity wave power spectrum by
\be\label{gravspec}
P_g^{1/2}=\sqrt{\frac{k^3}{2\pi^2}}\left|\frac{v_k}{a} \right|.
\ee
Again, deep inside the horizon ($aH/k \ll 1$) we have the free field solution
for a mode
\be\label{gravinside}
v_k\propto\frac{1}{\sqrt{2k}}e^{-ik\tau}
\ee
which freezes after leaving the horizon ($aH/k\gg 1$), $v_k\propto a$.

%%%%%%%%%%%%%%%%%%%%%%%%%%%%%%%%%%%%

\subsection{Tensor and scalar modes for a generic fluid}

In order to see the effect of the second phase of inflation on the modes
that have been generated and left the horizon during the first phase,
we rewrite the equations in terms of physical properties
of the cosmic fluid, i.e. instead of considering the scalar field and its
potential, $\phi$ and $V(\phi)$, we express Eq. (\ref{uequ})
in terms of the equation of state of the fluid, $\w$.
This allows us to study general fluids and their effect on
the primordial power spectrum without the need to specify the potential.

Here we concentrate on fluids with $\w > -1$,
but a similar analysis is straightforward to carry out also when $\w \leq -1$.

By defining the effective equation of state $p= \w \rho$ and using the
definitions of pressure and energy density for a scalar field with a
canonical kinetic term, Eq. (\ref{uequ}) can be written as
\begin{widetext}
\be\label{uequ3}
u_k'' - \frac12 (1+3\w)u_k'
+ \left[ \left(\frac{k}{aH} \right)^2
    -\frac 12(1-3\w)-\frac 34\frac{1-\w}{1+\w}\w'-
\frac 12 \frac{\w''}{1+\w}
+ \frac 14 \left(\frac{\w'}{1+\w}\right)^2\right] u_k = 0,
\ee
\end{widetext}
where $'\equiv d/d(ln(a))$. This is a general result valid for
$\w > -1$. In particular one can use Eq. (\ref{uequ3}) to study
the evolution of perturbations during transition periods in the early
universe when the effective equation of state changes.

Similarly for gravity waves, we have
\be\label{vequ1}
v_k'' - \frac12 (1+3\w)v_k'
+ \left[ \left(\frac{k}{aH} \right)^2
    -\frac 12(1-3\w)\right] v_k = 0.
\ee
Note the qualitative difference between the two equations, as long
as the equation of state is constant, both $u_k$ and $u_k$ will evolve
in a similar fashion. However, when the equation of state changes, i.e.
when inflation stops (or a second phase of inflation starts) , scalar and 
tensor modes will evolve differently.

%%%%%%%%%%%%%%%%%%%%%%%%%%%%%%%%%%%%%%%

\section{Modes skimming through the second phase}

In order to solve the mode equations numerically, one needs
to specify the initial conditions. We adopt the approach also
followed, for example, in~\cite{andrew,richard} and start the
calculation for each mode deep inside the
horizon, where it can be accurately described by the
free field solution
\be \label{freefield}
u_k,\ v_k=\frac{1}{\sqrt{2k}}\e^{-ik\tau}.
\ee
By integrating Eq.~(\ref{uequ3}) and Eq.~(\ref{vequ1}) for each $k$-mode,
one can determine the produced spectrum via Eqs. (\ref{power},\ref{gravspec}).
Due to the linearity of the equations, in practice it is useful to integrate the real
and imaginary parts of $u_k$ and $v_k$ separately.

\begin{figure}
\includegraphics[width=7cm]{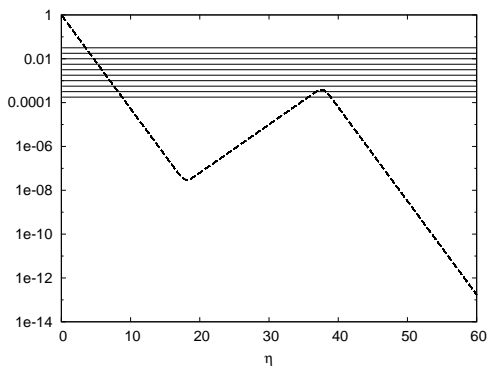}
\caption{Different $k$-modes and $(\ln(aH))^{-1}$ as a function of $\ln(a)$,
$\Delta\eta=3/2$.}\label{kandH}
\end{figure}

The equation of state changes from inflation, with $\w=\w_0=-0.99$, to a
matter dominated universe with $\w=\w_1=0$ and then back to an inflationary
epoch. The transition between the different phases is modeled
with a hyperbolic tangent,
\bea
\label{hyptan}
\w(\eta) & = & \w_0+\frac 12(\w_1-\w_0)\Big(\tanh\big(\Delta\eta\left(\eta-\eta_0\right)\big)+
\nonumber\\
& &
\tanh\big(\Delta\eta\left(\eta_1-\eta\right)\big)\Big),
\eea
where $\eta_0$ marks the end of the first phase of inflation and the
beginning of the matter/radiation dominated epoch, $\eta_1$ is when the second
inflationary phase begins and $\Delta\eta$ is a parameter determining
the smoothness of the transition.

In a realistic situation we expect that the transition from
matter domination to MSSM inflation be smooth. The MSSM
flat direction has frozen near the saddle point and carries a
constant potential energy $V_{\rm MSSM}$.

The oscillations of the $\chi$ fields which drive assisted inflation, see
Eq.~(\ref{1}), behave as matter and dominate the universe during the
intermediate stage. Their amplitude is redshifted by the Hubble
expansion until the MSSM flat direction takes over. As a result, since
there are no sudden changes in the potential of assisted or MSSM inflatons,
the equation of state of the universe changes smoothly from that of
matter domination to inflation.

The transition to the second phase of inflation can be accurately modeled as follows. 
Consider a universe filled with two fluids,
$\rho_{1,2}$ with two different equations of state $\w_{1,2}$ so that
$\rho_{1,2}=\rho_{1,2}^0\exp(-3(1+\w_{1,2})\eta)$. The effective
equation of state of the two fluid system is given by
\bea\label{wtwofluids}
w & = & \frac{p_1+p_2}{\rho_1+\rho_2}=\frac{\w_1\rho_1+\w_2\rho_2}{\rho_1+\rho_2}
\nonumber \\
&=& \frac{\w_1\rho_1^0\e^{-3(1+\w_1)\eta}+\w_2\rho_2^0\e^{-3(1+\w_2)\eta}}
{\rho_1^0\e^{-3(1+\w_1)\eta}+\rho_2^0\e^{-3(1+\w_2)\eta}}\nonumber\\
& = & \frac{(\w_2-\w_1)}{2}\tanh \left(\frac 32(\w_2-\w_1)\eta-
\frac 12 \ln\frac{\rho_2^0}{\rho_1^0} \right)\nonumber\\
& & +\frac 12 (\w_1+\w_2).
\eea
Hence, for a matter dominated universe which later enters an inflationary phase
dominated by an almost constant vacuum energy density, we see that $\Delta\eta=3/2$.

In numerical calculations we will select  $\eta_0=18$, $\eta_1=38$ and
the numerical calculation is run until $\eta=60$.

%%%%%%%%%%%%%%%%%%%%%%%%%

%%%%%%%%%%%%%%%%%%%%%%%%%%

\begin{figure}
\includegraphics[width=7cm]{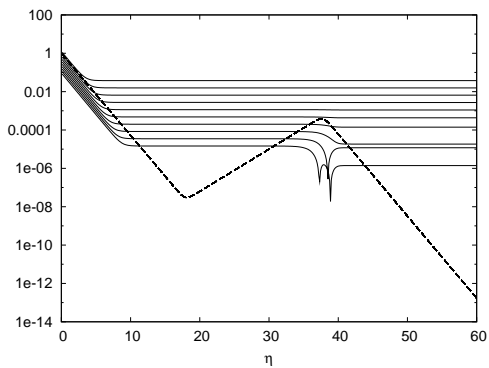}
\caption{The amplitude of the scalar modes
(solid line), $|u_k/z|$, along with $(\ln(aH))^{-1}$ (dotted line) for
$\Delta\eta=3/2$ is plotted. Normalization is chosen arbitrary. }
\label{spectra0}
\end{figure}

\begin{figure}
\includegraphics[width=7cm]{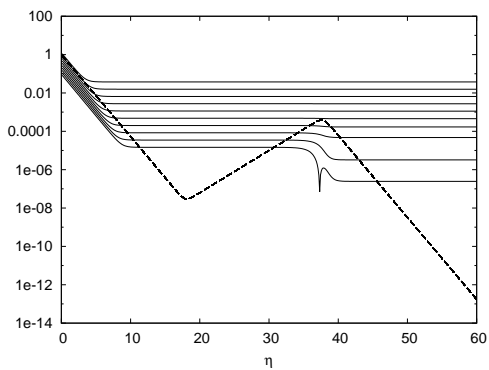}
\caption{The amplitude of the tensor modes (solid lines), $|v_k/a|$
along with $(\ln(aH))^{-1}$ (dotted line) is plotted, for a smooth transition from
matter domination to (second phase of) inflation, $\Delta \eta=3/2$. Normalization
is chosen arbitrary. }
\label{spectra-1}
\end{figure}

\begin{figure}
\includegraphics[width=7cm]{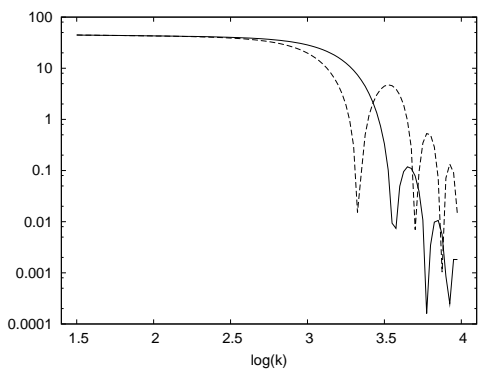}
\caption{The scalar (dotted line) and tensor (solid line)
spectrum for $\Delta\eta=3/2$ are plotted. Normalization is arbitrary and the mode
that just grazes the horizon as the second phase of inflation
starts has $\log_{10}(k)\approx 3.47$. One can clearly see that the scalar and
tensor modes that re-enter the horizon before the second phase of inflation
start decaying.}
\label{spectra}
\end{figure}

%%%%%%%%%%%%%%%%%%%%%%%%%%%%%%%%%

\subsection{Evolution of scalar modes}

In Fig.~(\ref{kandH}) we show $(\ln(aH))^{-1}$ along with a number of
modes with different $k$ as a function of $\eta$. A given mode exits/enters the
horizon when $k=aH$, i.e. when the lines cross. Modes start deep inside
the horizon, then freeze as they leave the horizon. As is clear from the figure,
some of the modes then re-enter during the matter dominated epoch separating
the inflationary periods. The modes shown here are sample of the modes
used in calculating the resulting spectrum, i.e. we have used more modes.

In Fig. (\ref{spectra0}) we have shown the evolution of
the amplitude of the scalar modes, $|u_k/z|$, as a function of $\eta$.
We see that as the modes leave the horizon,
they are frozen so that $u_k\propto z$, as expected. Modes that later
re-enter the horizon get re-processed. Modes that are just entering
the horizon as the second phase of inflation begins are re-processed and
even modes that never re-enter the horizon undergo some
evolution~\footnote{During the transition it is conceivable to amplify
the non-Gaussian parameter, $f_{NL}$, following Refs.~\cite{NG}. We
will come back to this issue in a separate publication.}.

%\footnote{We have also studied
%sharper transitions and confirmed that the modes evolve as expected:
%a sharper transition implies larger derivatives $\w',\ \w''$ and hence,
%according to Eq. (\ref{uequ3}), larger deviations from a scale invariant
%spectrum due to non-adiabatic evolution of the vacuum.}.

%%%%%%%%%%%%%%%%%%%%%%%%%%%%%%%%%%%%%

\subsection{Evolution of tensor modes and distinctive features}

In Fig. (\ref{spectra-1}) we have shown the evolution of
the amplitude of the tensor modes, $|v_k/z|$, as a function of $\eta$.
The tensor spectrum on the very large angular scales is solely
given by the one arising from the assisted inflation.

The mode that just grazes the horizon, $k=k_c$, as the second inflationary
phase begins, can be calculated by considering the evolution of
$\ln(1/(aH))$. During the first inflationary phase $H_{\rm assist}$ is constant, and
hence $\ln(1/(aH_{\rm assist}))=-\eta-\ln (H_{\rm assist})$. After the universe becomes
matter/radiation dominated, $\ln(1/(aH))=(1+3\w)\eta/2-\ln(H_{\rm assist})$, where
$\w$ is the equation of state of the fluid dominating the universe between the
inflationary phases. In the numerical calculation we have set, $H_{\rm assist}=1$,
$\w=0$ so that when MSSM inflation starts at $\eta=38$, $\ln(1/(aH))=(38-18)/2-18=-8$.
Since a mode leaves the horizon when $k_{c}=aH$, $k_c=\exp(8)$ or $\log_{10}(k_c)\approx 3.47$.

\begin{figure}
\includegraphics[width=7cm]{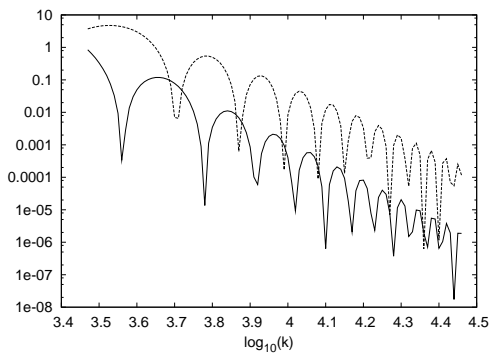}
\caption{Details of the tensor (solid lines) and scalar (dotted line) spectrum 
for $\Delta\eta=3/2$. The largest mode corresponds to the critical mode grazing the horizon.}
\label{specdet}
\end{figure}

In Figs. (\ref{spectra},~\ref{specdet}) we show the tensor and scalar power spectrum, 
${\cal P}_{\cal R}(k)$ and $\mathcal{P}_g(k)$ respectively.
From the figure one can read that both spectra are dropping with $k$ with a large power. In the case of
tensor perturbations, i.e. $\mathcal{P}_g(k)\sim (k/k_c)^{-6}$ in good agreement with
analytical calculations \cite{Brandenberger}. The rate of decay is constant over the studied range of $k$, hence a general prediction of the scenario described here is that the tensor spectrum is very red with distinct bumps. The remnant scalar spectrum from the first phase of inflation also drops rapidly with increasing $k$. Furthermore, the tensor and scalar spectra exhibit anti-correlation so that an increase in scalar power means less power in the tensor power at the same scale (and vice versa).

%%%%%%%%%%%%%%%%%%%%%%%
%%%%%%%%%%%%%%%%%%%%%%%

%%%%%%%%%%%%%%%%%%%%%%%

If the whole observed scalar spectrum is due to the MSSM inflation and does not have features
due to remnants from the assisted inflationary phase, then one expects that the tensor spectrum
should be observable only at the very largest scales. For example, if the tensor-to-scalar ratio
at $l(\log_{10}(k_{c})=3.47)=2$ is of the order $0.8$, then at $l=10$, corresponding to 
$\log_{10}(k)=4.47$ in Fig.~\ref{specdet}), the tensor-to-scalar ratio is roughly $\sim 10^{-4}$. 

So far we have assumed that
the current horizon corresponds to the critical mode $k_c$. 
If we relax this assumption along with the details of the first inflationary phase, we have much more freedom in the scenario. For example, if the amplitude of scalar perturbations from the first phase of inflation is larger than the observed value, implying a higher scale for the assisted inflationary phase, one can have the current horizon corresponding to a mode that has been sufficiently re-processed in between inflationary phases. Then one would still have the second phase of inflation producing scalar perturbations on all observable scales but now the horizon size can correspond to a mode which is smaller than the critical one, i.e. $\log_{10}(k)\approx 3.7$. For this mode one can read from 
Fig.~(\ref{specdet}) that the tensor-to-scalar ratio is amplified by an order of magnitude from its initial value. The other option is to consider scenarios where the observed scalar spectrum is a combination of the two inflationary phases. This naturally implies that there is a break in the scalar spectrum, therefore one would expect to see feature in the scalar spectrum. In all of the cases the main observable prediction remains, a very red tensor spectrum with distinct bumps observable at the largest scales. This can be associated with anti-correlated features in the scalar spectrum.

The anti-correlation between scalar and tensor spectra can be qualitatively understood as follows. 
First of all note that the frequencies of tensor and scalar modes are different, see Eqs.~(\ref{uequ3},\ref{vequ1}), they differ by the time variation in the equation of state as it changes from matter domination to inflation. This mainly affects modes whose wavelength is of the order of horizon size when the second phase of inflation began, i.e. modes to the left of Fig.~(\ref{specdet}). As a result the tensor and scale modes start oscillating at different times which leads to a phase difference between them. For modes which are deep inside the horizon at that time, i.e. modes to the right of Fig.~(\ref{specdet}), $(k/aH)$ is large and the first term inside the brackets in Eqs.~(\ref{uequ3},\ref{vequ1}) dominate, leading to the same frequency for scalar and tensor modes. Hence the modes tend to become correlated to each other at larger values of $k$ (larger values of multipoles, $l$). These are the features which distinguish our scenario with respect to the case where gravity waves arise {\it solely} from a single phase of inflation.

%%%%%%%%%%%%%%%%%%%%%%%%%%

\section{Conclusion}

We have illustrated a simple example where we can generate
observable gravity waves from assisted inflation at high scales
$H_{\rm assist} \approx 10^{13}$ GeV, while the origin of
temperature anisotropy arises mainly from a low scale
MSSM inflation $H_{\rm MSSM} \sim 1 $ GeV. The latter is also responsible for 
reheating the universe to the SM baryons and the cold dark matter.

The two phases of inflation are separated by a matter dominated
epoch from coherent oscillations of the fields that drove assisted
inflation. Being absolute gauge singlets, as usually assumed,
these fields do not directly couple to matter, therefore, it is conceivable
that they do not decay before the second phase of inflation kicks in.

The assisted inflation phase also creates the appropriate initial
conditions for the MSSM inflation. This is a natural outcome due
to the high scale of inflation, which imparts large quantum fluctuations
to the MSSM flat direction that push it to the vicinity of the desired VEV,
$\phi_0 \sim 10^{14}$~GeV, within $\sim 80$ e-foldings of assisted inflation.

The tensor-to-scalar ratio for a mode which is skimming through the
second phase of inflation is bounded by Eq.~(\ref{imp}).
For $H_{\rm assist} \approx 10^{13}$~GeV and $H_{\rm MSSM} \sim 1$~GeV
the observed tensor-to-scalar ratio is given by $r_{\rm observed} \leq 0.8$, observed only
at the largest angular scales. The tensor perturbations have a distinctive feature
shown in Figs.~(\ref{spectra}, \ref{specdet}), with a power spectrum that falls as $\sim k^{-6}$.
This is in sharp contrast to the spectrum of gravity waves generated by a single phase of inflation
alone, where such a re-processing of tensor modes would never occur.
The future experiments may be able to distinguish between the two scenarios.

%%%%%%%%%%%%%%%%%%%%%%%%%%%%%%%%

A.M is partly supported by the European Union through Marie
Curie Research and Training Network ``UNIVERSENET'' (MRTN-CT-2006-035863) and
by STFC (PPARC) Grant PP/D000394/1. He also acknowledges the support
from the Niels Bohr Institute during the course of this work. T.M is supported
by the Academy of Finland. We would also like to thank the referee for making
critical comments which has helped to improve our presentation.

%%%%%%%%%%%%%%%%%%%%%%%%%%%%

\end{document}